# ADVANCEMENTS IN CROWD-MONITORING SYSTEM: A COMPREHENSIVE ANALYSIS OF SYSTEMATIC APPROACHES AND AUTOMATION ALGORITHMS: STATE-OF-THE-ART

Mohammed Ameen[1] and Richard Stone[2]

[1]Department of Human-Computer Interaction, Iowa State University, USA, Ames
[2] Industrial and Manufacturing Systems Engineering Department, Iowa State University, USA, Ames

### ABSTRACT

*Growing apprehensions surrounding public safety have captured the attention of numerous governments and security agencies across the globe. These entities are increasingly acknowledging the imperative need for reliable and secure crowd-monitoring systems to address these concerns. Effectively managing human gatherings necessitates proactive measures to prevent unforeseen events or complications, ensuring a safe and well-coordinated environment. The scarcity of research focusing on crowd monitoring systems and their security implications has given rise to a burgeoning area of investigation, exploring potential approaches to safeguard human congregations effectively. Crowd monitoring systems depend on a bifurcated approach, encompassing vision-based and non-vision-based technologies. An in-depth analysis of these two methodologies will be conducted in this research. The efficacy of these approaches is contingent upon the specific environment and temporal context in which they are deployed, as they each offer distinct advantages. This paper endeavors to present an in-depth analysis of the recent incorporation of artificial intelligence (AI) algorithms and models into automated systems, emphasizing their contemporary applications and effectiveness in various contexts.*

## 1. INTRODUCTION

Crowd monitoring is the process of evaluating and watching how big crowds of people behave and move about in public places. Crowd monitoring is now essential for guaranteeing public safety, security, and effective crowd management due to the rise in popularity of major gatherings and events [1]. Crowds in various places such as streets, parks, airports, classrooms, swimming pools, and innumerable other locations are typically monitored by Close Circuit Television cameras (CCTV)[2]. The disadvantages of CCTV cameras are limited area coverage, high power consumption, installation problems, mobility, and constant monitoring by operators [3]. Many algorithms and approaches have been developed to efficiently identify, track, and evaluate crowd behavior and density to monitor and control crowds[4], [5]. These algorithms can interpret data from various sources, including sensors, social media feeds, and CCTV cameras, and they can also deliver real-time forecasts and insights[6]. Machine learning, deep learning, and computer vision algorithms are important methods used in crowd surveillance[7]. The accuracy and dependability of machine learning models in dynamic and complex situations, such as crowds, is a challenge, as well as crowd behavior, evolves over time, models may need to be periodically retrained to maintain accuracy; otherwise, their efficiency will gradually decrease[8]. Although machine learning and artificial intelligence have advanced, human input is still essential for live crowd monitoring for decision-making[9]. The complexity and wide range of variables that might affect crowd behavior make it difficult for algorithms to generate reliable forecasts. Here are various justifications for why live crowd monitoring requires human participation to have real-time monitoring, contextual understanding, and decision-making. In general, crowd monitoring systems and their operators confront several difficulties, and it is crucial to have knowledgeable staff and reliable mechanisms in place to handle and address these difficulties successfully. The main challenge of the crowd-monitoring approach is to establish a prediction model that evaluates both the role of operators and machine learning algorithms in crowd-monitoring in terms of decision-making.

## 2. CROWD MONITORING SYSTEMS (CMSS)

As the global population continues to grow, monitoring and managing crowd dynamics has become essential to maintaining public safety and ensuring efficient resource allocation. A crowd monitoring system is a cutting-edge technological solution that combines various data sources, such as computer vision, sensor networks, and social media data, to better understand the crowd. By leveraging a proper framework equipped with appropriate technologies, CMSs can process vast amounts of data, enabling authorities to make informed decisions, prevent dangerous situations, and optimize crowd-management strategies [10]. CMSs are classified into two main types vision-based(image) and non-vision-based (signal data) technology, which has significantly improved the accuracy and efficiency of CMSs, particularly in detecting and tracking individuals within large gatherings [11]. Initially, the crowd data is directly/ indirectly acquired using devices that are vision and non-vision based[12]. Monitoring individuals or large gathering by using these technologies require three main aspects which are crowd counting, localization, and behavior, as shown in Figure 1 [3].

Despite these advances, several challenges remain in developing and deploying CMSs, such as ensuring data privacy, addressing computational complexities, and overcoming the limitations of individual data sources. As research in this domain continues, it is vital to develop novel techniques and methodologies to address these challenges, as well as to maintain a multidisciplinary approach that encompasses the integration of emerging technologies.

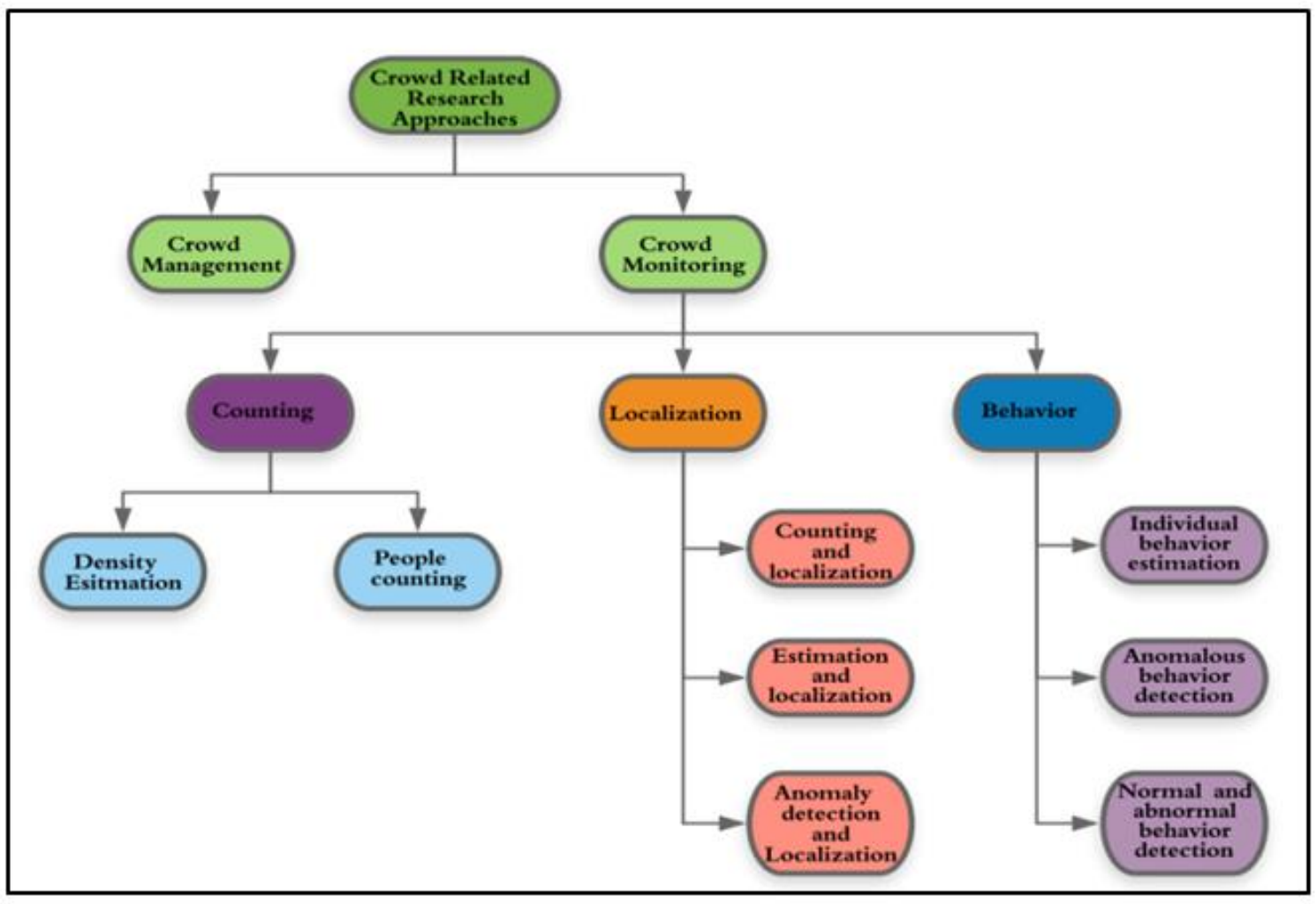

Figure 1. Crowd-related research approaches.

## 2.1. Vision-based

Vision-based methods utilize video cameras to identify human actions and gestures from video sequences or visual data. Vision-based methods employ cameras to detect and recognize activities using different computer vision techniques, such as object segmentation, feature extraction, feature representation, etc.[13] . Vision-based methods have recently become a primary approach for crowd monitoring systems, like hotspots, streets, and buildings because Vision-based systems make excellent use of the infrastructure provided by security networks [14] . This method is based on several computer vision concepts that help in recognizing activities which are:

- **Object segmentation:** is one such technique, partitioning a digital image into multiple segments to help isolate individuals within a crowd[15]. Additional techniques include edge detection, which identifies the boundaries of objects within an image, aiding in segmenting different individuals within a crowd [16].
- **Feature extraction and representation**: complements this by identifying specific 'features' or 'attributes' within the data, such as facial features or body shapes, helping distinguish different individuals [17], [18]. Feature representation translates these attributes into a format that computer systems can understand and process[19].
- **Pattern recognition**: plays a crucial role by identifying certain patterns within the data, such as movement patterns, to offer insights into crowd behavior[20]. Image recognition is employed to identify specific objects or features within an image, such as identifying individuals or detecting signs of overcrowding [21].

These techniques provide a comprehensive understanding of crowd dynamics, enabling real-time monitoring and managing large crowds during events. The deployment of vision-based systems during mass gatherings has been integral to maintaining the safety of the attendees. The systems effectively utilize the infrastructure provided by security networks, such as CCTV cameras and other surveillance equipment, for optimal use of available resources. These tools provide real-time data, which are then processed and analyzed using computer vision techniques to detect potential risks, manage crowd flows, and direct resources where they are most needed. These systems harness the power of sophisticated computer vision algorithms and machine learning techniques to process the captured data. For instance, Convolutional Neural Networks (CNNs) are often employed for image classification tasks, contributing significantly to the accuracy and efficiency of these systems [22]. Other algorithms include Support Vector Machines (SVM) and Decision Trees for pattern recognition and predictive modeling [23].

Security systems use computer vision algorithms and machine learning techniques to process the captured data and provide insights into crowd behavior, density, movement patterns, and potential risks[5]. They are useful for ensuring crowd safety and security during large gatherings, events, or public spaces. While security systems employing computer vision algorithms and machine learning techniques have significantly improved the surveillance landscape, the role of CCTV operators remains crucial in maintaining a comprehensive and effective surveillance system. Numerous studies have demonstrated the efficacy of diverse approaches for the automated identification of potential hazards, enabling operators to respond appropriately and make decisions, particularly in intricate scenarios[24]. Despite their merits, relying solely on fully automated mechanisms for threat detection may not be advisable, as they could potentially overlook genuinely perilous incidents or generate unwarranted false alarms[25]. Currently, most systems require, each operator must control the video sequences of multiple security cameras simultaneously[26]. This fact, along with inherent human characteristics such as fatigue or loss of attention over time, can lead to ignoring important events or responding too late, causing significant damage and making these video surveillance systems inefficient or useless[27].

## 2.2. Non-Vision-based

The field of non-vision-based techniques for crowd monitoring has not received as much attention as its vision-based [28]. A variety of these non-vision approaches have been employed for managing crowd dynamics during large-scale events. Some prevalent methods include Wi-Fi, Bluetooth, RFID, and cellular networks. In their work, Felemban, Emad A., et al. [29] discuss the application of non-vision-based technologies, such as Wi-Fi and RFID, in the context of the Hajj pilgrimage (refer to Figure 2). Furthermore, El-Adl, Gamal H. [30] introduced a smart bracelet that incorporates an array of non-vision technologies for use during the Hajj, including GPS, GIS, LED, Smart Card, wireless sensors, and their respective tags. Over recent years, several comprehensive literature reviews have been conducted in this domain.

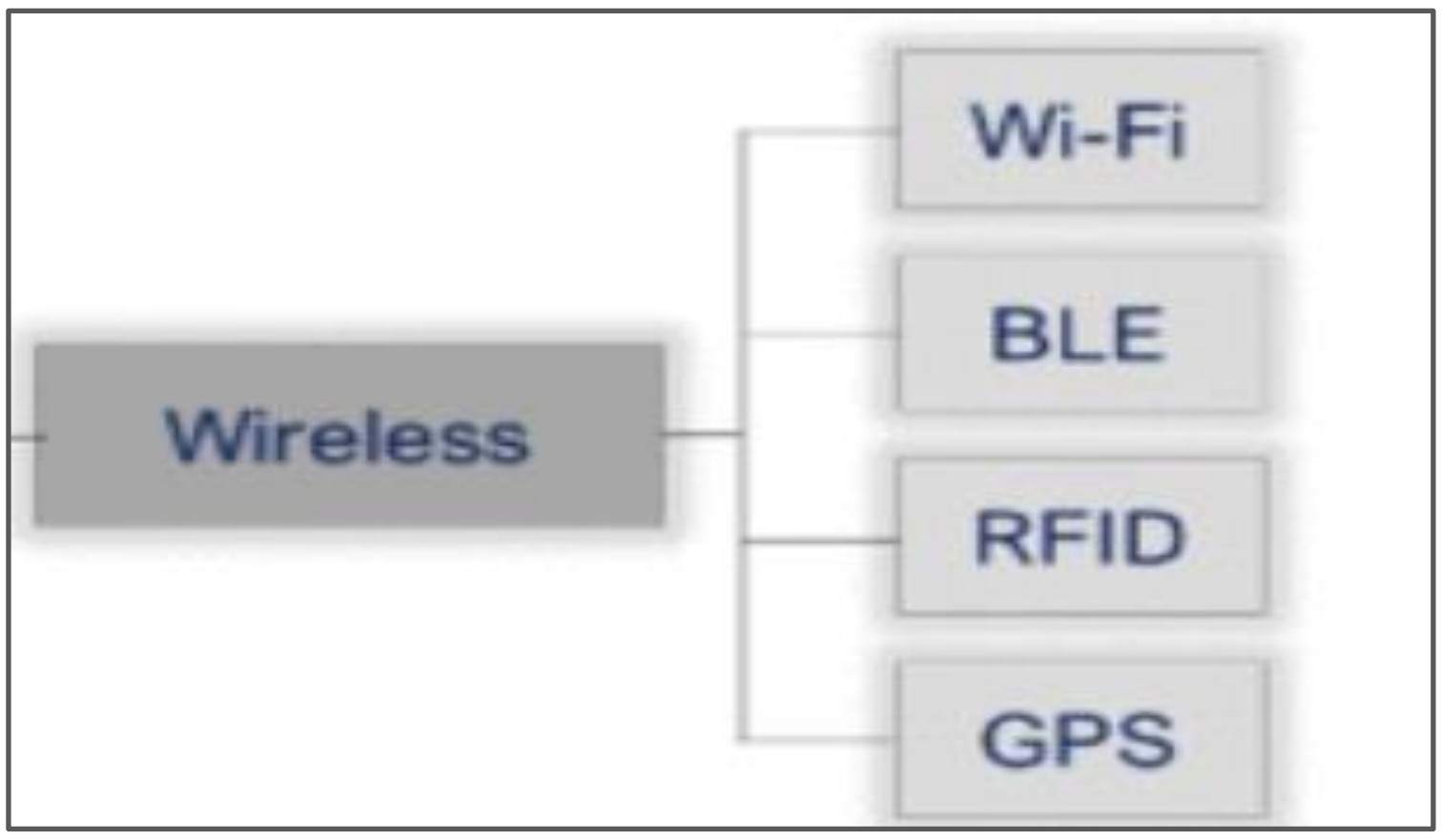

Figure 2. Non-vision Technologies.

In recent years, Wi-Fi technology has gained significant popularity for tracking the movement and location of Wi-Fi-enabled devices [12]. Similarly, Bluetooth technology utilizes a comparable technique, albeit with a more limited range compared to Wi-Fi [31]. Another study proposed a combination of Bluetooth and Wi-Fi technologies for tracking purposes, along with infrared sensors for counting individuals [32] . In contrast, RFID technology leverages radio waves to identify and track objects or individuals bearing specific tags [33].

RFID fundamentally comprises several integral elements, namely antennas, tags, and readers [205]. Each RFID tag is distinguishable via a unique identifier that can broadcast the information stored within. This retained data may either be exclusively readable or exhibit both read-write attributes. RFID tags can be systematically classified into distinct types, based on their unique features and energy sources. The operational functionality and power requirements substantially differentiate one type of RFID tag from another. Active tags are characterized by an integrated power supply, usually a battery, which autonomously powers the tag's operations [34], [35]. On the other hand, passive tags do not encompass an internal power source. Instead, they rely on the signal emitted by the base station. Another type, the semi-active or semi-passive tags, employs a hybrid approach. Similar to passive tags, they backscatter the carrier signal emanating from the base station. However, akin to active tags, they can be equipped with an inbuilt battery. The battery, in this case, does not directly contribute to the communication process but rather supplements the internal operations of the tag, enhancing overall performance [36]. However, Bluetooth Low Energy (BLE) technology function within the 2.4 GHz frequency band, a spectrum which they share with Wi-Fi transceivers, given the similar indoor propagation properties they possess [37] . Nevertheless, it's crucial to underscore that Bluetooth exhibits a significantly lower energy consumption profile when compared to Wi-Fi communication systems. In fact, Bluetooth's power requirements are an order of magnitude less than that of Wi-Fi, thereby significantly extending the battery life of the devices utilizing it [38].

Torkamandi et al. [39]employed an algorithm that utilized ephemeral MAC addresses for estimating crowd size. Their findings demonstrate that the method accurately estimates the number of devices across a diverse range of scenarios and environments. Furthermore, in their 2019 study, Groba et al. [40] utilized MAC addresses while also examining distance- and time-based filters. They proposed filtering by recognition to estimate the size and localization of a marching crowd. Satellite-based radio navigation technologies, such as the Global Positioning System (GPS), are extensively employed in open areas, particularly with a large number of mobile users. The widespread availability of GPS tracking devices have facilitated the collection of spatio-temporal information on human movements and behavior, providing new challenges and opportunities in social and economic research [41]. In human gatherings, primary uses are in tracking the movement of individuals within a crowd, providing valuable insights into crowd dynamics. GPS tracking can allow for the modeling of multiple crowd parameters, including crowd density, pressure, velocity, and turbulence, which are primarily visualized as heat maps[12]. Also, GPS-based mobility can provide a low-cost and remaining solution for crowd monitoring and tracking[42]. However, GPS is ineffective within indoor environments as it provides inaccurate results when used in confined spaces thus.

Ultimately, technologies that rely on vision have demonstrated a notable superiority in crowd monitoring compared to those that don't, primarily due to their capability to supply detailed and complex data about the behavior of crowds. Visual technologies, including computer vision algorithms and deep learning neural networks, surpass their non-visual equivalents in their ability to gather and scrutinize a broad spectrum of data more effectively.

This data includes crowd density, movement patterns, and behavioral analytics that are critical in managing and understanding crowd dynamics. As a result, vision-based systems can provide more accurate and timely responses, enhancing safety and crowd management efficacy. In contrast, non-vision based systems, offer less context-specific detail and can miss critical visual cues that can be crucial in ensuring effective crowd management. also, A major limitation of non-visual sensors is people unwilling to use them[43].

## 3. Algorithms in Surveillance Systems for Detecting Abnormal Behavior

Technology has become ubiquitous in daily activities, and people use it to ensure they have top-of-the-range surveillance systems. Analytics are useful in creating better surveillance while applying algorithms that enhance sensing and detection systems[44]. Machine and deep learning methods are integrated with video management systems (VMS) to augment detection solutions with artificial intelligence. Machine learning (ML) is vital for automating image processing and simplifying computer vision. Considering the progress in surveillance for different functions, this literature review synthesizes evidence of the various surveillance algorithms for abnormal behavior detection.

Video monitoring specialists employ advanced computational tools to scrutinize visual and auditory data captured by surveillance cameras to accurately identify individuals, objects, and specific scenarios[45], [46]. These scenarios may encompass atypical behaviors, which indicate deviations from established norms within restricted zones. In the past, traditional surveillance systems relied on human operators to vigilantly observe video feeds[47], [48]. However, contemporary algorithms have significantly improved surveillance capabilities by detecting anomalous behavior and promptly issuing real-time alerts[49]. The rules-based approach is the most cost-effective and widely adopted technology in the surveillance domain but may not perform optimally in highly dynamic settings, such as hospitals and educational campuses. It is primarily due to the challenges in effectively separating a rule's behavior from its environmental context and the intricacies of its implementation mechanism, as the approach merely stacks rules and accounts for exceptions [50]. There are several factors such as object size, motion, and velocity are taken into account while devising these surveillance algorithms[51]. Consequently, the rules-based approach's limitations in active environments highlight the indispensability of artificial intelligence in augmenting surveillance efficacy.

Active environments are hard to survey, as it is difficult to set rules that effectively discriminate between normal and abnormal behavior[52]. To improve surveillance monitoring, artificial intelligence can be employed in various ways, such as automating the monitoring process, identifying anomalies, and including real-time alerts [53]. It also classifies them as typical human behavior with various features, such as orientation [54]. Therefore, the classification of patterns in data, including in surveillance monitoring systems, can be done with the use of strong techniques like machine learning, and deep learning which will help to instantly evaluate vast amounts of data and find patterns or abnormalities that might point to security threats or other significant events.

### 3.1. Generative Adversarial Network (GAN)

GNA is a deep learning-based approach that has gained wide acceptance as it consistently achieves higher performance compared to traditional approaches, especially in detecting abnormal activities[55]. The research on algorithms and how they enhance abnormal behavior detection have been increasing in the last few years, GAN algorithm has received a lot of attention in anomaly detection research because of its special capacity to generate new data[56], [57]. Also, GNA algorithm produces an ideal image with super-resolution reconstruction, data enhancement, and anomaly detection[58]. There are different factors that play a role in the success of GAN algorithm which are[57]:

1. GANs algorithm demonstrates desirable performance by executing an unsupervised learning method.
2. The GAN's generative model can be utilized to successfully increase the dataset that is currently available, resolving the issue of reduced robustness brought on by a lack of data.
3. GAN is consisting of a generator and a discriminator which follows the procedure of confrontation learning.

Alafif et al. studied normal and abnormal detention in an active environment, Hajj, the Islamic pilgrimage event attended by millions, and found that GAN was pertinent in abnormal behavior accuracy detection of 79.63% in large crowds[59]. Ultimately, the goal is to mitigate abnormal behavior, such as collisions and crowd flow accidents. Notably, GAN produces data all at once and not pixel-by-pixel which presents better results than other models [1]. GANs have various applications, such as object detection, high-quality image generation, and inpainting. Other applications include super-resolution images, person re-identification, and video prediction[44]. Thus, GANs can survey large crowds to identify abnormal behaviors.

### 3.2. Convolution Neural Network (CNN)

Convolution neural network Algorithms can test deviation from typical human behavior by assessing suspicious events when they occur such as distinguishing patterns of behavior that might indicate criminal activities[60]–[62]. The researchers combined a CNN, which directly surveys visual patterns, with the long-short memory model (LSTM) thus, this combination provides a high rate of classification accuracy and precision, sensitivity, and specificity at 93.8%, 91.6%, and 98.6% respectively[63]. CNN can also use various applications that are ideal for recognition, such as facial recognition, biometric authentication, etc.[53], [64]. However, the main one disadvantage of CNN is that it requires training of large datasets thus, it will be a costly and time-consuming procedure[65]. In order to attain high levels of accuracy and resilience in the network's predictions of CNN, a Pre-trained model that has been trained on substantial amounts of training data may occasionally be utilized as the foundation for new applications, which can assist to decrease the quantity of training data needed[66].

### 3.3. Long Short Term Memory Model (LSTM)

LSTM is a recurrent neural network (RNN) that has been effective since it can analyze sequences of data in addition to single data points, which is advantageous for anomaly identification[67]. The following justify their use and suitability for categorizing, analyzing, and generating predictions based on time series data[68]:

1. Traditional RNNs frequently experience vanishing and expanding gradient issues, and LSTM effectively handles these issues.
2. LSTM has an advantage over traditional RNNs because it's relatively insensitive to the duration of gaps between time series data.
3. LSTM Gating mechanism allows for efficient management of long-term memories.

LSTM is fundamentally like RNN, but it can utilize a mechanism called a " gate " to learn about long-term dependency [69]. Both neural networks process chronological data but they differ since RNNs remember data from previous inputs [53]. However, LSTM is more dynamic as it can remember previous inputs and achieve long-term dependencies[70]. With long-term dependencies, LSTM is more beneficial as it does not suffer the vanishing gradients problem, unlike RNN, which is difficult to train[71]. The primary limitation of LSTM is the need for training large datasets, which can be a laborious and costly process[53]. Since LSTM can be used for recognition and classification, it uses various applications such as speech recognition and video analysis[72]. LSTM processes and classifies chronological data with long-term dependencies. Overall, LSTM achieves high accuracy in training models, especially in recognizing and classifying abnormal behaviors, such as fighting in surveillance.

### 3.4. Gaussian Mixture Model (GMM)

As highlighted in the previous research on surveillance algorithms, machine learning can effectively assess activity at regional functions, prisons, airports, and other density-filled places with tremendous activity every second. While extending the analysis of algorithm models in surveillance, some researchers have tested whether algorithms can separate background and foreground as clutter and complex occlusions make noisy backgrounds[73]. The algorithm model adopted to separate the foreground from the background was the Gaussian mixture model (GMM), as it can evaluate scene changes[74], [75]. GMM provides more contextual information than other algorithm models, which can be helpful when handling ambiguous data points[76]. However, the GMM has certain disadvantages, including high false alarm rates, the need for augmented calculations to detect different rates, and increased computational costs[77]. These reasons could be justified because GMM models have problems categorizing features, assuming normal distributions, and making assumptions about cluster shapes. More importantly, the augmented calculations illustrate the need for sufficient data for each cluster[78]. However,

Ghasemi and Ravi demonstrate that when a GMM model is adapted to reduce limitations, it shows high detection accuracy and processing speed when evaluating crowded spaces[77].

### 3.5. The Support Vector Machine (SVM)

The support vector machine (SVM) model tests anomalies in forestry and environmental applications. The SVM can detect the loss of individual trees or evaluate surface temperatures[79]. SVM models are also used to assess human behaviors and classify these actions, such as kicking, punching, hitting, and fighting[80]. SVM can evaluate different human behaviors and accurately detects them at 72% and 68% in two datasets (Manjula and Lakshmi 4). Accordingly, SVM offers various benefits such as it is memory efficient and working in case of a class margin separation, which is ideal for real-time surveillance and related applications[81]. SVM models can be combined with others, including the Smoothed Total Variation (STV) that help to detect as they denoise and generate final classification maps[82]. In isolation, SVM is an efficient method of detecting abnormal behaviors as it achieves high performance within the computed time[70]. However, the training speed of an SVM is slow when dealing with huge amounts of data[83]. Several studies combine two models when testing inconsistencies in human behavior. For example, Sridhar et al. noted that combining models allows for high accuracy in detecting human behavior for various activities such as killing or fighting[84]. Thus, the advantage of combing two algorithm models is to increase detection and prediction and identify anomalies occurring within online monitoring applications.

### 3.6. Random Forest (RF)

Random Forest (RF) is an algorithm encompassing a random forest classifier algorithm comprising various decision trees[85]. One advantage of RF is that it improves accuracy and precision by using decision trees, which can reduce the overfitting of datasets[86]. Using decision trees means that researchers can identify a target value, such as fighting with two or more branches (subsets) that represent feature values for various aspects such as facial expressions during an argument[87]. In their study, Evans et al incorporated contexts such as Christmas, Easter, football, half marathon, and public holidays in the decision tree to assess if context impacted the false alerts in a Traffic Management Centre[88]. The research found that RF led to fewer false alerts by 25%, while the accuracy rate increased by 95% in a crowded traffic area[88]. Alafif also used an algorithm that combined CNN with the RF algorithm, resulting in the evaluation of small-scale crowd abnormal behaviors[59]. Also, Alafif found that both datasets had 97% and 93.71% accuracy rates when assessing small crowds. However, if accuracy is the goal, increasing trees should be a priority, which can be laborious and slow the model. Regardless, RF can be applied to predict customer behavior, such as demand and fraud, or for diagnosing in healthcare. Therefore, RF uses implies that it is ideal for classifying and identifying regression problems.

## 4. Accuracy Rates of Different Models and Algorithms in Violence Detection

The literature review has illustrated that algorithm models are critical in violence prevention by increasing accuracy in detection. Thus, one way to evaluate studies is to assess whether they ensure the effectiveness of surveillance algorithms in testing the level of accuracy in crowded areas and uncrowded areas. The accuracy rates differ when testing crowded or uncrowded places using machine learning, and deep learning techniques. CNN method resulted in 78.64% accuracy in both crowded and uncrowded areas of investigation[89]. They were more accurate in using violence detection techniques to identify abnormal human behavior. Equally, some deep learning models have a higher accuracy rate in crowded places[90]. When applying CNN models in deep learning, Mu et al. found they were 90% accurate in detecting violence in crowded areas[81]. SVM and GMM proved more effective for crowded places, achieving an accuracy of 82-89%[91]. Thus, this analysis demonstrates that surveillance methods present varying results when testing violence in crowded and uncrowded areas.

Table 1. Advantages and disadvantages of the machine and deep learning algorithms

| **Machine and Deep learning algorithm** | | |
|---|---|---|
| **Algorithm** | **Advantages** | **Disadvantages** |
| GAN[92] | Generate the sharpest image sample | It is difficult to optimize because of the unstable training dynamic. |
| CNN[93] | Feature extract and learning | Require huge amount of data for training |
| LSTM[93] | Having the capacity to store several layers of information or data and being more adept at identifying and utilizing long-range context | The task of sequence transcription is not supported; input sequences must be aligned with target sequences. |
| GMM[94] | • Has flexibility in terms of cluster covariance.<br>• Its computational requirement is inexpensive.<br>• It is used as a model with hidden parameters. | • It is not accurately model for quickly varying background<br>• Require enough memory.<br>• Time-consuming in terms of initializing the parameter in the background model |
| SVM[89] | • limit the risk of error<br>• good algorithm to model non-linear relations<br>• comprehensive/robust | • Training tiresomely slow<br>• Hazard of overfitting<br>• The model can be difficult to understand |
| RF[89] | • Easier to tune than GBM<br>• Harder to overfit | • A large number of trees could result in slow performance for real-time prediction. |

## 5. Open Challenges and Future Trends

The particular context and type of the data at hand will influence the selection of the most appropriate algorithm for identifying suspicious activities. CNNs are frequently employed compared to alternative algorithms; however, concerns regarding training data persist. Integrating multiple algorithms can enhance detection performance by capitalizing on the advantages of each technique and mitigating their respective shortcomings. This fusion of methods can yield more precise and consistent predictions. Singh et al. employed CNN, RNN, and LSTM, achieving an impressive 97% accuracy in anomaly detection[95].

Numerous researchers have endeavored to create automated monitoring systems, yet they have encountered challenges, and the performance has not yet met the desired standards. No system has been developed that achieves 100% detection accuracy and a 0% false detection rate, particularly in the context of videos featuring complex backgrounds[96].

This underscores the continued necessity for human expertise. Despite the integration of advanced artificial intelligence technologies within surveillance systems, the human element remains indispensable. Automated surveillance systems operate under the supervision of human operators, who possess the innate ability to adapt to various situations and respond accordingly. The discernment of human experts is crucial for comprehending situational factors that may impact an individual's behavior and subsequently evaluating the validity of AI-generated alerts.

These professionals possess a unique skill set, which includes keen attention to detail, exceptional observational abilities, and the capacity to make swift, informed decisions. They often undergo specialized training to hone their aptitude for recognizing suspicious activities, analyzing patterns, and understanding context-specific cues.

Future work, the objective should be to devise more potent systems that harness sophisticated deep learning algorithms, which can yield accurate outcomes in demanding situations, thereby bolstering the model's comprehensive performance. It is essential to thoroughly examine the interplay between the surveillance system operator and their ability to contribute to the system's efficacy. There is a notable absence of methodologies for measuring the efficacy of hybrid security systems (CCTV operator and automated technologies), impeding comprehensive evaluation of these systems' performance. The recommended future work is to establish new strategies that help to evaluate the performance of CCTV operators with artificial intelligent algorithms in the context of detecting abnormal behavior by using several approaches. For instance, measuring the operator's performance, cognitive load, trust, and confidence while monitoring human activities with an automated system.

## 6. CONCLUSION

We present a comparative analysis of various crowd-monitoring system methodologies and the employment of machine learning and deep learning algorithms for the identification of anomalous behaviors. To facilitate this comparison, we have carefully chosen a selection of cutting-edge models that have gained significant prominence in the field of crowd-monitoring systems.

In this study, we offer a thorough comparative examination of the two approaches, namely vision-based and non-vision-based crowd monitoring systems, with a keen focus on their respective strengths and limitations. We contend that vision-based systems are poised to supersede their non-vision-based counterparts, owing to the intrinsic benefits they provide, such as access to a richer data source, the capacity for enhanced analytical capabilities, and greater adaptability across diverse scenarios.

The abundant data yielded by vision-based systems enables the efficient application of machine learning and deep learning methodologies, encompassing techniques such as crowd analysis, object detection, and speech recognition[95].

Research demonstrates that cutting-edge machine learning and deep learning algorithms are consistently surpassing expectations and delivering exceptional outcomes within the realm of vision-based crowd monitoring systems. these advanced algorithms enables the extraction of intricate high-level features and the identification of complex patterns within the available data, ultimately resulting in heightened accuracy and refinement in crowd-monitoring applications.(Table1) presents a comprehensive summary of the pairwise comparisons utilizing ML and DL techniques that have been applied within the past five years. These methods present an array of promising prospects for augmenting the scope and bolstering efficacy in the field. Furthermore, they can be harnessed for a diverse assortment of additional applications.

## AUTHORS

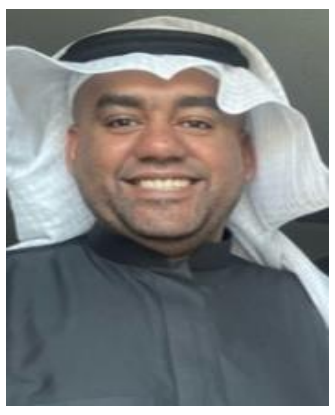

**Mohammed Ameen**, a Ph.D. student in the Department of Human-Computer Interaction at Iowa State University. I received my master's degree in computer science and engineering from University of Connecticut in the United state in 2017. I received my bachelor's degree in Information technology from King Abdul Aziz University in 2012. I worked at King Abdul Aziz University as a lecturer from 2017 to 2020 in the information system department. My current research focuses primarily on computer vision and human behaviours.

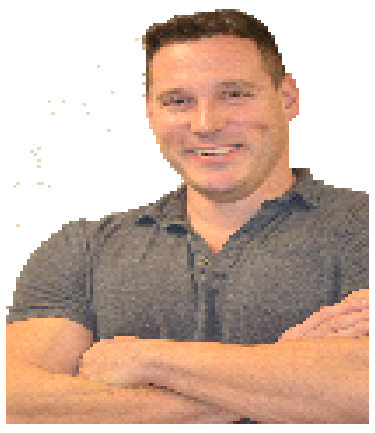

**Richard T. Stone**, Ph.D. is an Associate Professor in the Department of Industrial and Manufacturing Systems Engineering at Iowa State University. He received his Ph.D. in Industrial Engineering from the State University of New York at Buffalo in 2008. He also has an MS in Information Technology, a BS in Management Information Systems as well as university certificates in Robotics and Environmental Management Science. His current research focuses primarily on the area of human performance engineering, particularly applied biomedical, biomechanical, and cognitive engineering. Dr. Stone focuses on the human aspect of work across a wide range of domains (from welding to surgical operations and many things in between). Dr. Stone works extensively in tool, exoskeleton, and telerobotics technologies, as well as classic ergonomic evaluation.